# Multiferroicity and magnetoelectric coupling in α-$CaCr_2O_4$

Kiran Singh[1*], Charles Simon[1] and Pierre Toledano[2]

[1] *Laboratoire CRISMAT, CNRS UMR 6508, ENSICAEN, 6 Bd. du Maréchal Juin, 14050 Caen Cedex, France*

[2] *Laboratory of Physics of Complex Systems, University of Picardie, 33 rue Saint-Leu, 80000 Amiens, France*

## Abstract:

Ferroelectricity in the incommensurate helical magnetic phase (below $T_N$, 43K) of alpha (α) $CaCr_2O_4$ has been confirmed by pyroelectric measurements. Magnetoelectric and magnetodielectric coupling exist below $T_N$ and are proportional to the square of magnetic field. From symmetry analysis, we suggest that the presence of an external electric field destabilizes the symmetrical 2221' phase and stabilizes 21' symmetry. This provides a unique system in which polarization varies as the *fourth-degree* of the order-parameter amplitude and exhibit a vanishingly small value below the first-order transition at $T_N$, as observed experimentally.

PACS: 75.85.+t, 77.22.-d, 77.22.Ej, 77.80.B-

*Corresponding author: Kiran Singh
e-mail: kiran.singh@ensicaen.fr; kpatyal@gmail.com

**Introduction:**

In these days, the fascinating interplay between ferroelectric and magnetic order parameter makes multiferroic materials as one of the interesting topics of condensed matter physics. These materials have attracted immense attention due to strong magnetoelectric coupling and several potential technological applications [1]. The search of such materials has been initiated after the discovery of magnetoelectric coupling in $TbMnO_3$ [1]. Such materials are very rare [2] and on the basis of their microscopic origin these materials are classified into several groups [3]. Recently, different *"frustrated magnets"* exhibit magnetoelectric coupling: e.g. $RMn_2O_5$ [4], $Ni_3V_2O_8$ [5], $MnWO_4$ [6], $Ba_{0.5}Sr_{1.5}Zn_2F_{12}O_{22}$ [7], $CoCr_2O_4$ [8], $ABX_2$ (A=Cu, Ag; B=Fe, Cr and X=O and S) [9, 10, 11, 12, 13]. In all these system the ferroelectricity has been observed in the incommensurate magnetic phase.

Among frustrated magnets, the triangular antiferromagnetic lattice has been studied extensively especially delafossites. The crystal structure of $\alpha$-$CaCr_2O_4$ has been reported long time ago by Pausch et al. [14]. It crystallizes in an orthorhombic layered structure with triangular sheets of $CrO_2$ separated by $Ca^{+2}$ and mostly resembles to delafossites with a small distortion. Recently, the magnetic properties of $\alpha$-$CaCr_2O_4$ has been investigated in details by Chapon et al. [15]. It exhibits a long range antiferromagnetic incommensurate helical magnetic phase below $T_N$=43K having magnetic propagation wave vector q=(0,0.3317(2),0). Neutron diffraction data shows that the plane of rotation of spins is perpendicular to the wave vector. In materials with incommensurate phases, the lattice modulation is generally connected with a spatially varying electric polarization [16] (e.g. $Rb_2ZnCl_4$, $(NH_4)_2BeF_4$ and $K_2SeO_4$). At low temperature, these materials exhibit a first-order phase transition to a ferroelectric phase. Similarly, ferroelectricity has been reported in $TbMnO_3$ [1], $YBaCuFeO_5$ [17] and $CuFeO_2$ [9] in the incommensurate magnetic phase. Considering these observations, we could expect the field induced ferroelectricity in $\alpha$-$CaCr_2O_4$ in the incommensurate phase below $T_N$. Although the magnetic point group 2221' does not allow ferroelectricity, however, there is no report on the dielectric and ferroelectric studies of this system. In this context, it is worth to investigate dielectric and magnetoelectric properties of this system in details. In this paper, we report for the first time the temperature and magnetic field dependent dielectric and ferroelectric properties of $\alpha$-$CaCr_2O_4$. A clear dielectric peak observed at $T_N$ in association with incommensurate magnetic phase. Our pyroelectric results confirm that this ferroelectric (incommensurate magnetic phase) to paraelectric (paramagnetic phase) transition is of first order.

**Experimental:**

Polycrystalline $\alpha$-$CaCr_2O_4$ were prepared by spark plasma sintering method. The phase purity of the sample was studied by room temperature x-ray diffraction. DC magnetization measurement was performed in zero field cooling and field cooling mode at 0.3T magnetic field by using Quantum Design superconducting quantum interference device (SQUID). Clear antiferromagnetic transition observed at 43K (not shown here). Dielectric measurement was performed on a thin parallel plate capacitor. Silver paste was used to make electrodes. The dielectric and magnetodielectric studies were made by using Agilent 4284A LCR meter at four different frequencies (5 kHz – 100 kHz) and different magnetic field (0, 0.5 and 10T) during heating and cooling (1K/min). Isothermal magnetodielectric were performed at different temperatures between ±14T with sweep rate of 100Oe/sec at 100kHz. Ferroelectricity was confirmed from pyroelectric measurements using Keithley 6517A electrometer. Sample was cooled from room temperature to 55K without any electric field. A poling electric field of ±630kV/m was applied at 55K during cooling to align the electric dipoles and removed at 8K. Polarization vs time was recorded for 5000 sec to remove the stray charge (if any) before measuring the pyroelectric current. Magnetoelectric coupling was observed by measuring the polarization vs magnetic field at 42K. The sample was cooled with same electric field up to 42K in a similar way as mentioned for polarization *vs* temperature. Magnetic field was ramped between ±14T (100Oe/sec) many times to insure the reproducibility of the data. In each case magnetic field was applied perpendicular to the direction of electric field.

**Results and discussion:**

We have observed a lamda ($\lambda$) like peak in dielectric permittivity at $T_N$ (Fig. 1 (a)). This peak is frequency independent (not shown here). A small hysteresis is observed during cooling and warming measurements (not shown here), indicating the signature of first order transition. The losses are very small ($<10^{-4}$ below 250K). This anomaly is consistent with the anomaly observed in magnetization and heat capacity measurements [15], which indicates the coupling between charge and spin orders. The dielectric value decreases almost linearly from room temperature to 200K and then decreases slowly up to 90K. Below 90K dielectric permittivity again increases with decreasing temperature (inset of Fig. 1 (a)). The similar dielectric anomaly is expected for related materials of this family, e.g. $\alpha$-$SrCr_2O_4$ which has similar magnetic structure [18].

The existence of ferroelectricity in the incommensurate helical magnetic phase is proved by pyroelectric measurements (Fig. 1 (b)). Although the remnant value is very small, these results are highly reproducible and repeated many times. The direction of polarization flips by reversing the sign of poling electric field. The polarization decreases discontinuously at $T_N$ which is associated to the order of phase transition. For ferroelectric materials the order of transition can also be confirmed by pyroelectric measurement. In first order transition remnant polarization decreases discontinuously while in second order transition it decreases continuously from ferroelectric to paraelectric phase [19]. In this case polarization drops discontinuously to zero at $T_N$, such behavior has been observed at first order ferroelectric to paraelectric phase transition [20]. The first order nature of this transition is confirmed from our pyroelectric measurements and also mentioned by Chapon et al. [15]. The type of phase transition is different from delafossites and the saturated polarization is also small as compared to delafossites. This difference could be related to different magnetic point group. The remnant polarization is comparable with spinel $CoCr_2O_4$ [8].

Recently Dutton et al. [18] investigated magnetic and structural properties of $\alpha$-$SrCr_2O_4$ which is analogous to $\alpha$-$CaCr_2O_4$. They have studied the variation of lattice parameters with temperature and observed an inflection point at $T_N$ in all lattice parameters without any change in symmetry and proposed spin driven structural distortion. They also proposed the possibility of small domains with monoclinic or triclinic symmetry. However, there is no report on temperature dependent lattice parameters of $\alpha$-$CaCr_2O_4$. High resolution diffraction measurement could be useful to state the lowering of the symmetry to monoclinic or triclinic symmetry which could be favorable for ferroelectricity.

Dielectric permittivity was also measured under different magnetic field to see if $T_N$ varies under magnetic field or not. There is no shift in dielectric anomaly when measured under high magnetic fields (up to 10T) (Fig. 2), this shows the robust nature of antiferromagnetic interactions in this sample. To confirm the magnetoelectric coupling, we have measured isothermal polarization vs time by ramping magnetic field up to ±14T (100Oe/sec) at 42K (Fig. 3). In this figure, we have presented the observed data, remnant polarization and background signal. We have extracted the average value of polarization signal by assuming symmetric behavior while increasing and decreasing magnetic field. Background is the difference between observed data and average value of polarization. Such analysis is useful to see the clear effect of magnetic field even for very small variation of polarization and also reported for delafossites $ACrO_2$ (A=Cu and Ag) [21]. The magnitude of remnant polarization at 42K is taken from polarization vs temperature measurement. From Fig. 3, one can see that polarization follows magnetic field (maximum at minimum field and minimum at maximum field) which again supports the presence of magnetoelectric coupling in this sample. The magnetic field dependence of polarization at 42K and dielectric permittivity at 10K is shown in Fig. 4 (a) and (b), respectively. Fig. 4 demonstrates that both polarization and dielectric permittivity decreases with increasing magnetic field and proportional of square of magnetic field. On the basis of symmetry analysis it is proposed that in this material polarization can appears under magnetic field [15]. On the contrary we observed electric polarization with out any external magnetic field and it decreases quadratically with magnetic field. To clarify this issue the studies on high quality single crystal is required.

On the basis of symmetry analysis two different explanations can be proposed for the observation of ferroelectricity in $\alpha - CaCr_2O_4$ below $T_N = 43$ K, which are consistent with the orthorhombic symmetry 2221' assumed for this phase by Chapon et al. [15] in the absence of applied electric field, and a vanishingly small value of polarization ($P < 1\mu C/m^2$) found under applied electric field of 630 kV/m. The preceding authors show that two incommensurate bi-dimensional order-parameters $(\eta_1, \eta_1^*)$ and $(\eta_2, \eta_2^*)$ are associated with the transition at $T_N$. The lowest symmetry phase induced by their coupling (Fig. 6 in Ref. 15) is a polar phase of monoclinic symmetry 21', which allows a spontaneous polarization $P_y$. One can therefore assume that a large poling electric field destabilizes the non-polar 2221' phase and induces a crossover to the ferroelectric phase.

The dielectric contribution to the transition free-energy, given by Eq. (1) in Ref. 15, reads as $\frac{P_y^2}{2\varepsilon_{yy}^0} + \delta P_y(\eta_1^2\eta_2^{*2} - \eta_1^{*2}\eta_2^2)$, where $\delta$ is a coupling constant and $\varepsilon_{yy}^0$ is the dielectric permittivity in the paramagnetic phase.

Putting $\eta_1 = \rho_1 e^{i\varphi_1}, \eta_1^* = \rho_1 e^{-i\varphi_1}, \eta_2 = \rho_2 e^{i\varphi_2}, \eta_2^* = \rho_2 e^{-i\varphi_2}$ the equilibrium polarization below $T_N$ varies as:

$$P_y^e = -\delta\varepsilon_{yy}^0 \rho_1^2 \rho_2^2 \sin 2(\varphi_1 - \varphi_2) \qquad (1)$$

Where $\varphi_1 - \varphi_2$ is the dephasing between the coupled order-parameters, which is arbitrary in the ferroelectric phase [15]. Thus, $P_y$ varies as the *fourth-degree* of the order-parameter amplitude and should exhibit a vanishingly small value below the first-order transition at $T_N$, as observed experimentally.

Another possible explanation of the emergence of a weak polarization below $T_N$ consists of assuming that the high electric field *decouples* the two order-parameters and *locks* the incommensurate magnetic wave-vector $\vec{k} = (0,0.3317,0)$ to the closest commensurate value $\vec{k} = (0, \frac{1}{3}, 0)$. Taking into account the corresponding symmetries of the irreducible representations, denoted $\Delta_1$ and $\Delta_2$ in Ref. 15, one can show that the same transition free-energy is associated with $(\eta_1, \eta_1^*)$ or $(\eta_2, \eta_2^*)$, which is:

$$F = \frac{\alpha}{2}\rho^2 + \frac{\beta}{4}\rho^4 + \frac{\gamma_1}{6}\rho^6 + \frac{\gamma_2}{6}\rho^6 \cos 6\varphi \qquad (2)$$

Where $\rho = \rho_1$ or $\rho_2$, and $\varphi = \varphi_1$ or $\varphi_2$. Minimizing $F$ with respect to $\rho$ and $\varphi$ yields the following results for the decoupled order-parameters:

1) The $(\eta_1, \eta_1^*)$ order-parameter, transforming as $\Delta_1$, induces three possible stable commensurate phases having the respective magnetic symmetries $P2_1/b$ for $\cos 6\varphi = 1$, $Pma2$ for $\cos 6\varphi = -1$, and $Pm$ for $\cos 6\varphi \neq \pm 1$, involving a three-fold multiplication of the $b-$ lattice parameter. In the orthorhombic $Pma2$ the polarization is along the $z$ axis, varying as:

$$P_z^e = -\delta\varepsilon_{zz}^0 \rho^3 \sin 3\varphi \qquad (3)$$

Whereas in the monoclinic $Pm$ phase the polarization is located in the (y,z) plane with the same dependence cubic dependence on the order-parameter given by Eq. (3) for the $P_y$ and $P_z$ components, which corresponds to a weak value of the induced polarization;

2) The $(\eta_2, \eta_2^*)$ order-parameter, transforming as $\Delta_2$, gives rise to three possible commensurate phases having the symmetries $P2/b(\cos 6\varphi = 1), Pmn2_1(\cos 6\varphi = -1)$ or $Pb(\cos 6\varphi \neq \pm 1)$ with the same three-fold multiplication of the paramagnetic unit-cell along $b$. The polarization in the $Pmn2_1$ phases is along the x axis whereas it is in the (x,y) plane for the $Pb$ phase, with a similar dependence on $\rho$ and $\varphi$ given by Eq. (3).

The different interpretations given for the electric-field induced polarization in $\alpha - CaCr_2O_4$ correspond to different orientations for the polarization and to different (incommensurate or commensurate) structures for the polar phase. Dielectric and structural measurements on single crystal should therefore essential to confirm the most suitable direction for ferroelectricity. However, since no lock-in transition was observed in Ref. 15, thus first interpretation in term of polar phase of monoclinic symmetry 21' is the most probable.

**Conclusion:**

The multiferroicity in $\alpha$-$CaCr_2O_4$ polycrystalline sample has been investigated. Our results show the spin induced ferroelectricity in the incommensurate helical magnetic phase. These results also confirm that the transition at $T_N$ is of first order. Magnetoelectric and magnetodielectric coupling exists and follows quadratic dependence of magnetic field which is the classical dependence of such phase. The similar properties are expected for other related materials which opens the avenue to find new multiferroics materials.

**Acknowledgments**: We acknowledge L. Chapon and A. Maignan for fruitful discussions.

**Figure captions:**

**Fig. 1.** (Color online) Temperature profiles of (a) dielectric permittivity measured at 100 kHz during warming (1K/min); inset shows the over all temperature behavior (8K to 300K) and (b) remnant polarization after poling with positive (black line) and negative electric field (red line) of $\alpha$-$CaCr_2O_4$.

**Fig. 2.** (Color online) Time and magnetic field dependence polarization of $\alpha$-$CaCr_2O_4$ at 42K (±14T; 100Oe/sec); arrows indicate the respective y-axis. Upper panel shows the as observed polarization data, middle panel shows remnant polarization and lower panel presents the background signal (see text for detail).

**Fig. 3.** (Color online) Relative dielectric permittivity vs temperature under different magnetic field (0, 0.5 and 10T) at 100 kHz.

**Fig. 4.** (Color online) Magnetic field dependent polarization at 42K (a) and dielectric permittivity (b) at 10K with increasing and decreasing magnetic field (100Oe/sec). Open circles represent experimental points and red line indicates the $H^2$ fitting. Here $\Delta\varepsilon=\varepsilon(H)-\varepsilon(H=0)$.

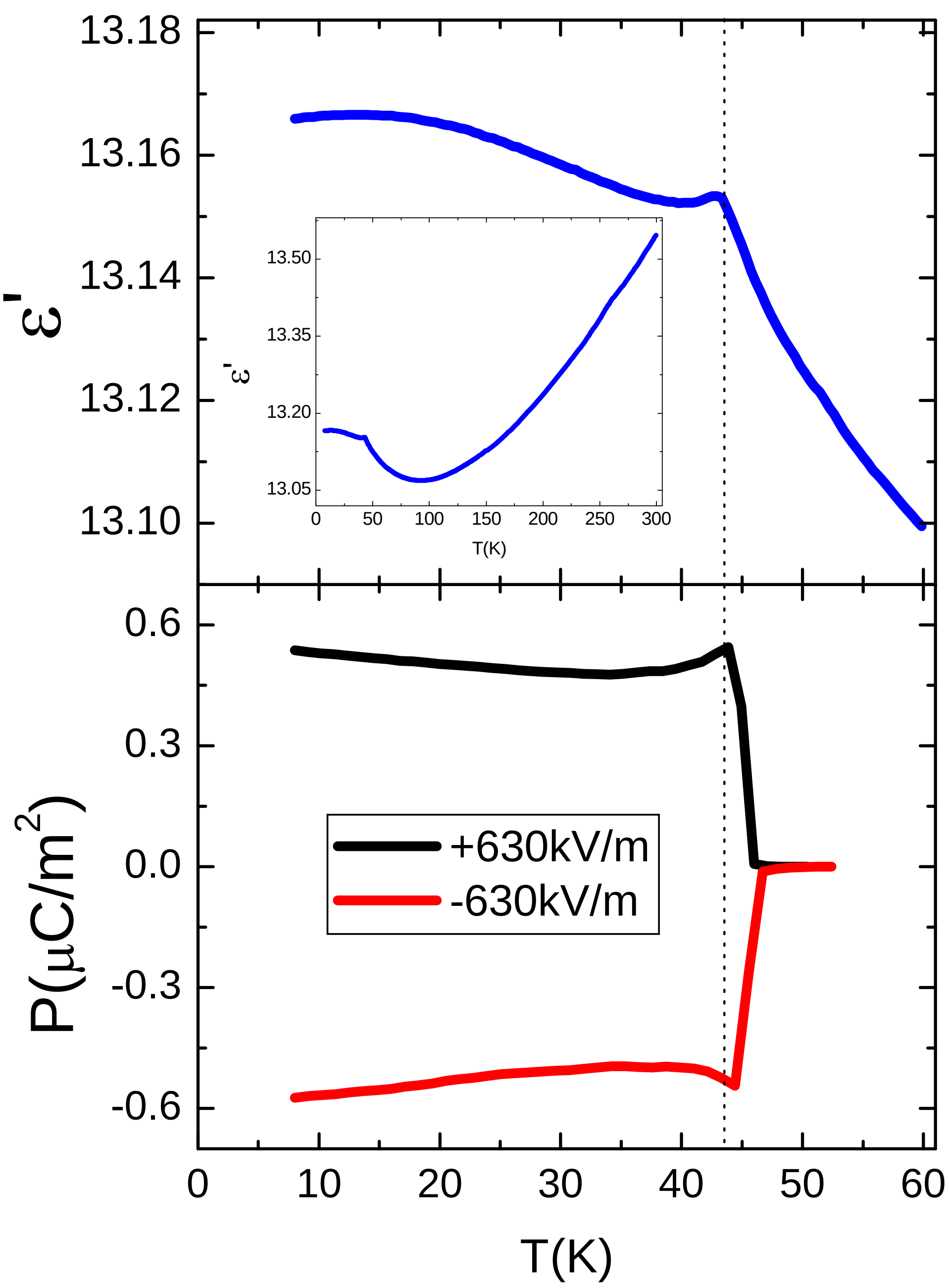

**Fig. 1.** (Color online) Temperature profiles of (a) dielectric permittivity measured at 100 kHz during warming (1K/min); inset shows the over all temperature behavior (8K to 300K) and (b) remnant polarization after poling with positive (black line) and negative electric field (red line) of α-$CaCr_2O_4$.

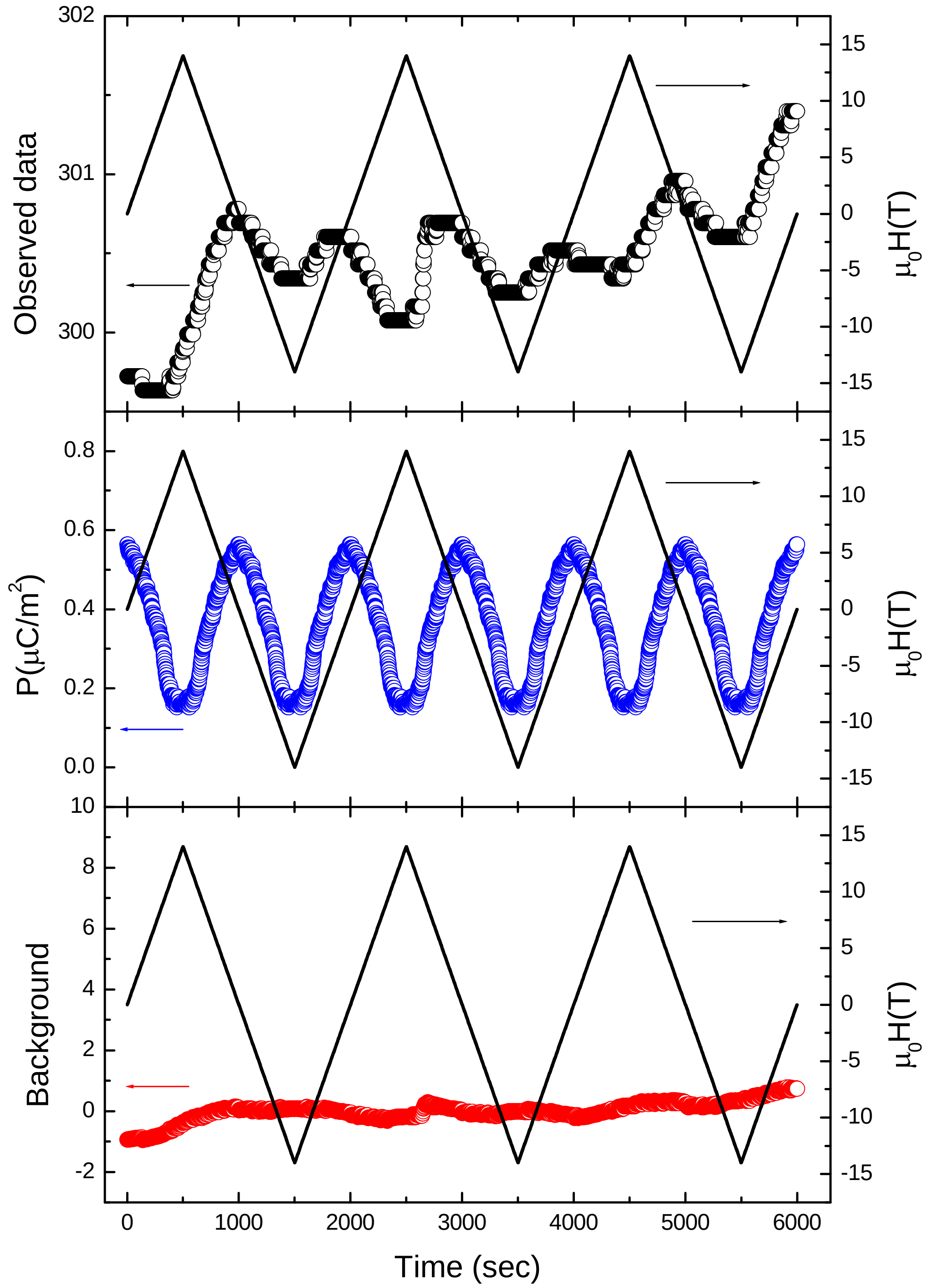

**Fig. 2.** (Color online) Time and magnetic field dependence polarization of α-$CaCr_2O_4$ at 42K (±14T; 100Oe/sec); arrows indicate the respective y-axis. Upper panel shows the as observed polarization data, middle panel shows remnant polarization and lower panel presents the background signal (see text for detail).

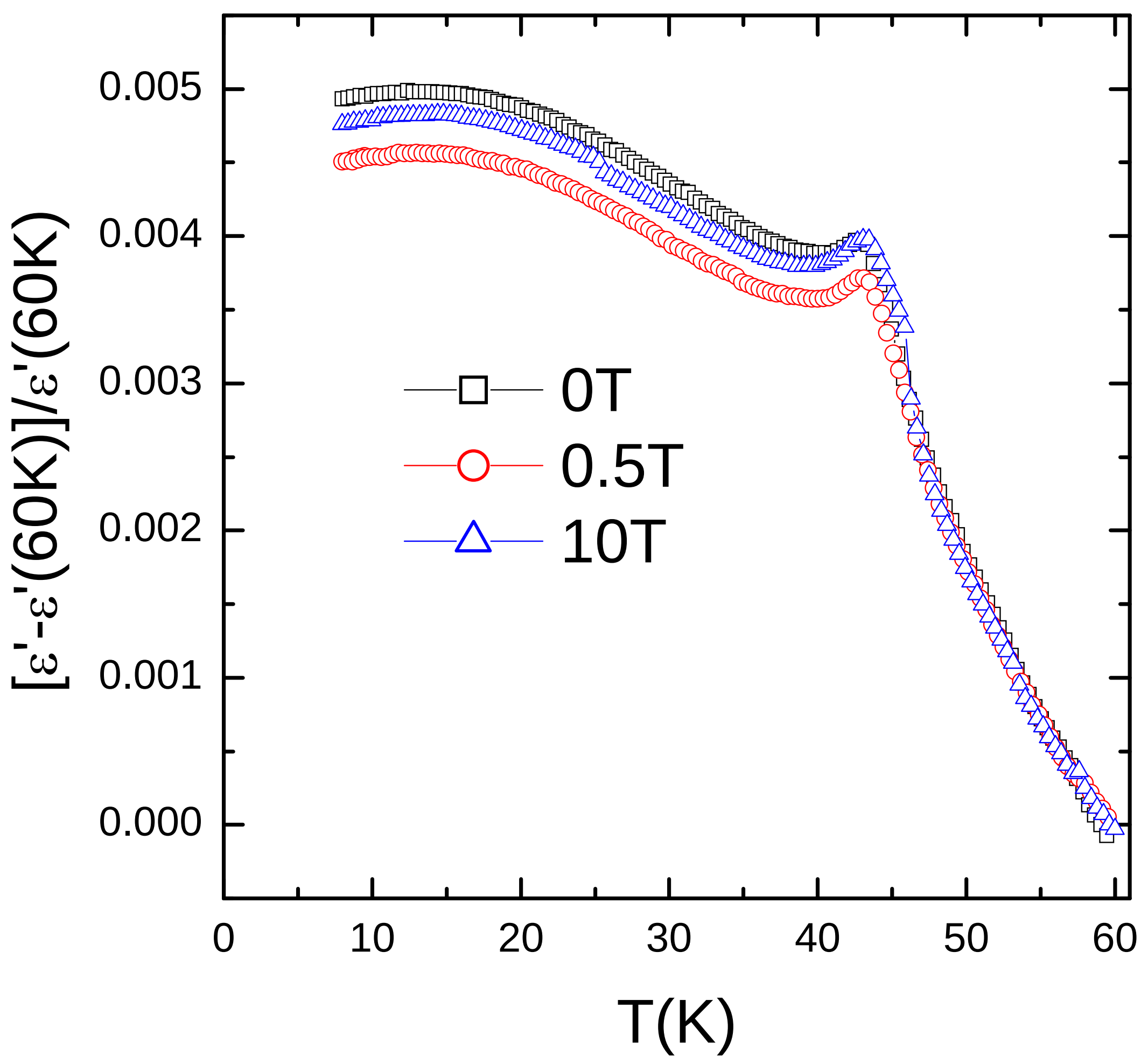

**Fig. 3.** (Color online) Relative dielectric permittivity vs temperature under different magnetic field (0, 0.5 and 10T) at 100 kHz.

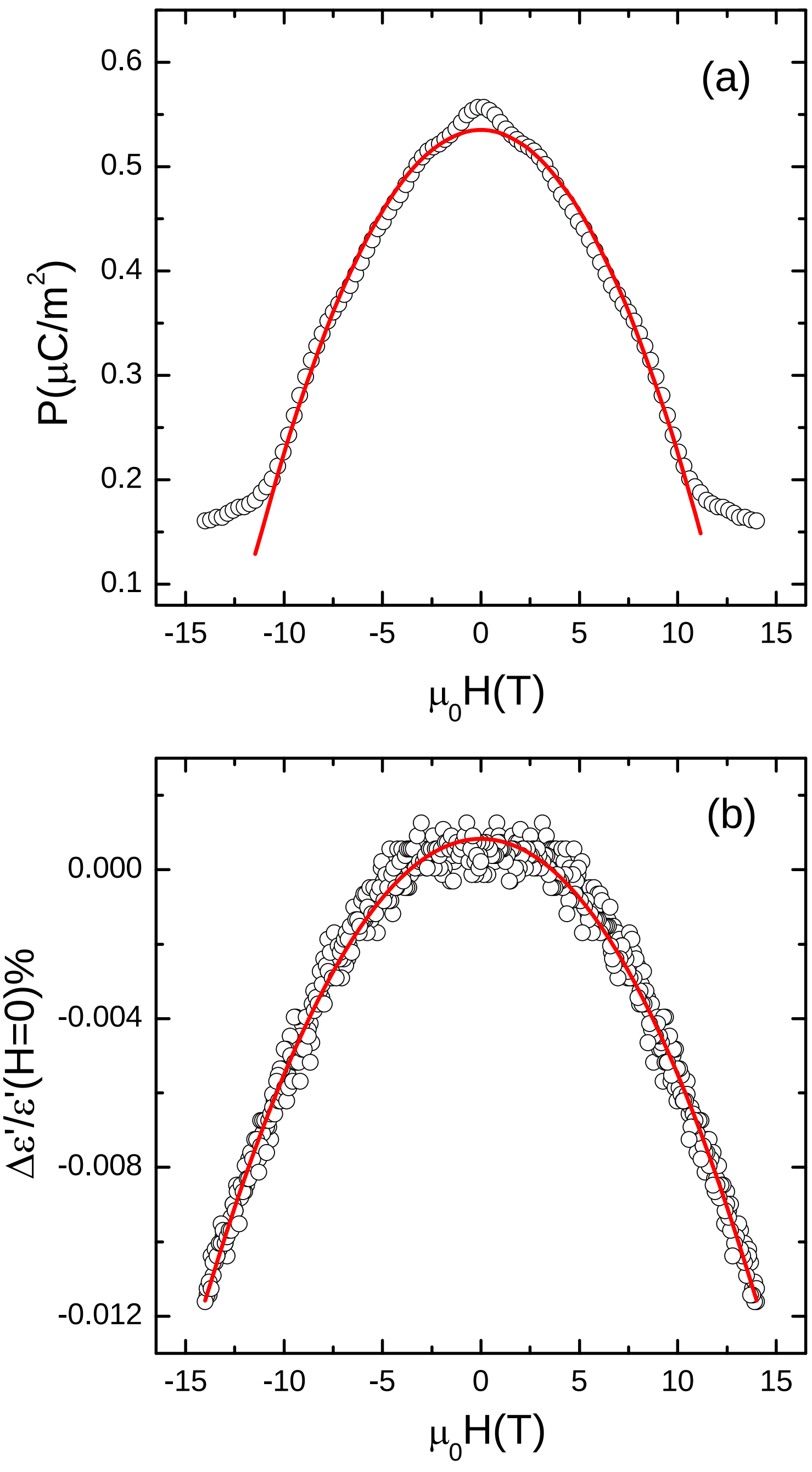

**Fig. 4.** (Color online) Magnetic field dependent polarization at 42K (a) and dielectric permittivity (b) at 10K with increasing and decreasing magnetic field (100Oe/sec). Open circles represent experimental points and red line indicates the $H^2$ fitting. Here $\Delta\varepsilon=\varepsilon(H)-\varepsilon(H=0)$.